\documentstyle[psfig]{mn}

\def\deg{^{\circ}}

\def\micron{$\mu m$}

\def\kms{\,km\,s$^{-1}$}

\def\radec{$\alpha,\delta$}
\def\xy{$x,y$}
\def\1pos{$x_1,y_1$}
\def\2pos{$x_2,y_2$}
\def\3pos{$x_3,y_3$}
\def\hb{H$\beta$}

\def\xieta{$\xi,\eta$}
\def\radec{$\alpha,\delta$}
\def\cent{$\alpha_{\rm{c}},\delta_{\rm{c}}$}

\def\sig68{$\sigma_{\rm{68}}$}

\title[Semi-automated Spectral Extraction]
{Semi-automated Extraction of Digital Objective Prism Spectra}
\author[C.A.L. Bailer-Jones et~al.]
{Coryn A.L.\ Bailer-Jones$^1$\thanks{Present Address: 
Mullard Radio Astronomy Observatory,
Cavendish Laboratory, Madingley Road, Cambridge, 
CB3 0HE, UK}\thanks{email: calj@mrao.cam.ac.uk},
Mike Irwin$^2$, Ted von Hippel$^3$\\
$^1$ Institute of Astronomy, Madingley Road, Cambridge, CB3 0HA, UK\\
$^2$ Royal Greenwich Observatory, Madingley Road, Cambridge, CB3 0EZ, UK\\
$^3$ Department of Astronomy, University of Wisconsin, Madison, WI 53706, USA\\
}
\date{Submitted 7 September 1997}
\pubyear{1998}

\begin{document}

\maketitle

\begin{abstract}
We describe a method for the extraction of spectra from high
dispersion objective prism plates. Our method is a catalogue driven
plate solution approach, making use of the Right Ascension and
Declination coordinates for the target objects.  In contrast to
existing methods of photographic plate reduction, we digitize the
entire plate and extract spectra off-line. This approach has the
advantages that it can be applied to CCD objective prism images, and
spectra can be re-extracted (or additional spectra extracted) without
having to re-scan the plate. After a brief initial interactive period,
the subsequent reduction procedure is completely automatic, resulting
in fully-reduced, wavelength justified spectra.  We also discuss a
method of removing stellar continua using a combination of non-linear
filtering algorithms.

The method described is used to extract over 12,000 spectra from a set
of 92 objective prism plates. These spectra are used in an associated
project to develop automated spectral classifiers based on neural
networks.
\end{abstract}



\section{Introduction}

The MK classification of stellar spectra
(Morgan, Keenan \& Kellman 1943\nocite{morgan_43a})
has been an important tool in the workshop of stellar and galactic
astronomers for more than a century. While improvements in
astrophysical hardware have
enabled the rapid observation of digital spectra, our
ability to efficiently analyze and classify spectra has not kept
pace. Traditional visual classification methods are clearly not
feasible for large spectral surveys. In response to this, we have been
working on a project to develop automated spectral classifiers
(von Hippel et~al.\ 1994; Bailer-Jones 1996; Bailer-Jones et~al.\
1997, 1998).  These classifiers, which are based on supervised
artificial neural networks, can rapidly classify large numbers of
digital spectra.

The development of these classification techniques has required a
large, representative set of previously classified spectra. The most
suitable data has been the spectra from the Michigan Spectral Survey
(Houk 1994)\nocite{houk_94a} and the accompanying MK spectral type and
luminosity class classifications listed in the {\it Michigan Henry
Draper} (MHD) catalogue (Houk \& Cowley 1975; Houk 1978, 1982; Houk \&
Smith-Moore 1988).
\nocite{houk_75a}\nocite{houk_78a}\nocite{houk_82a}\nocite{houk_88a}
This paper describes the data reduction techniques we developed to
extract and process these spectra.

\section{Plate Material}

The Michigan Spectral Survey was an objective prism survey of the
whole southern sky ($\delta < 12\deg $) from the Curtis Schmidt
Telescope at the Cerro Tololo Interamerican Observatory in Chile.  We
scanned a number of the plates from this survey 
using the APM facility in Cambridge
(Kibblewhite et~al.\ 1984)\nocite{kibblewhite_84a}.  This machine uses
a flying-spot laser and photomultiplier detector to digitize areas of
the plate.  The usual mode of use for prism plates
is to locate objects using their known co-ordinates
and then to scan just the region of interest, either by recording all
of the pixels or by parametrizing the object in real time (e.g.\
Hewett et~al.\ 1985\nocite{hewett_85a}). The
coordinates are often obtained from a direct
image of the same field taken on the same telescope.  Other
groups have also developed methods for the automated and
semi-automated extraction of prism spectra (e.g.\ Clowes, Cooke \&
Beard 1984; Flynn \& Morrison 1990; Hagen et al.\ 1995; Wisotzki et
al.\ 1996) often with the goal of identifying quasar spectra.

Our approach differs from the conventional method in the principal
respect that we used the APM in raster scanning mode to digitize the
entire plate. Subsequent plate reduction and extraction of the
spectra take place off-line.  The main reason for this approach is
that it can equally well be applied to CCD objective prism images,
which are increasingly replacing photographic plates.  Furthermore,
additional spectra can later be extracted very rapidly without
requiring access to a plate scanning machine.  Tests determined that
the optimal scanning resolution was 15 \micron, which corresponds to
1.45'' per pixel. While the site seeing is typically
better, the telescope has relatively poor tracking ability, and this led
to an effectively lower seeing (blurring). Table~\ref{plt_det} gives
details of the plates and the reduced spectra.  Figure~\ref{plate}
shows a typical plate.
\begin{table}
\begin{center}
\caption{Details of the plates and the extracted spectra.}
\begin{tabular}{|l|l|}\hline
Plate type      & IIaO \\
Plate size  & $\approx 20 \times 20 cm$ \\
            & $\approx 5\deg \times 5\deg$ \\
            & 12,000 $\times$ 12,000 pixels \\
            & 289 Mb (FITS) \\
Plate scale & 96.62 arcsec $mm^{-1}$ \\
Dispersion  & 108\,\AA/mm at H$\gamma$ \\
Scanning pixel size & 15$\mu$m \\
                    & $\Rightarrow$ 1.45 arcsec $pix^{-1}$ \\
                    & $\Rightarrow$ 1.6\,\AA\ $pix^{-1}$ at H$\gamma$\\
                    & (1.05\,\AA\,$pix^{-1}$ @ 3802\,\AA \\
                    & 2.84\,\AA\,$pix^{-1}$ @ 5186\,\AA)\\
Time to digitize one plate & 100 minutes \\
Coverage of final spectra & 3802--5186\,\AA \\
Magnitude limit of plates & B $\sim 12$ \\
Number of stars on 92 plates & $\approx 16,000$ \\ \hline
\end{tabular}
\label{plt_det}
\end{center}
\end{table}

\begin{figure}
\centerline{
}
\caption{(This figure is supplied as a separate JPEG file.)
  APM negative image of a 5$\deg$ objective prism plate from the
  Michigan Spectral Survey.}
\label{plate}
\end{figure}

As with the conventional APM method, we extract known objects on the
basis of their coordinates. However, due to the absence of any
appropriate direct plate material from which \xy\ coordinates could be
obtained, we used catalogue \radec\ coordinates of our target objects.
We discovered that the MHD \radec\ positions were
unreliable compared with those in the Positions and
Proper Motions (PPM) catalogue (R\"{o}ser \& Bastian
1991)\nocite{roeser_91a}, with an average discrepancy of
$\approx20''$. (The positions in the PPM South catalogue have mean random
errors of $0.1''$.) Hence where cross identifications between the MHD and
PPM catalogue entries were available (for about 85\% of the stars in
the MHD) we used the PPM coordinates.
Co-ordinates could of course be used from any other source catalogue.
Furthermore, because
the MHD is incomplete ($\sim 50\%$ of all stars down to $B \sim 11$)
we supplemented it with all PPM stars not listed in the MHD.
This supplement not only permits extraction of more spectra, but
helps us identify overlaps between neighbouring spectra.

\section{Image Reduction and Spectral Extraction}\label{data_red}

An objective prism disperses the light from every point in the field
of view, with the result that the spectra on the detector lack a
common wavelength zero point (Figure~\ref{plate}).  Thus the reduction
procedure must pay careful attention to the mutual wavelength
alignment (justification) of the spectra.  Another complication is
that because the plates were originally obtained for the purposes of
visual classification, they have been widened, thus increasing the
chance of overlaps between adjacent spectra. Finally, the \radec\
co-ordinates of the plate centres are only poorly known.

Given that we want to extract the spectra of objects with known
\radec\ co-ordinates, our reduction approach is to use a subset of
spectra to solve a parametrized mapping of the form \radec\ $
\Rightarrow $ \xy, and then to use this to obtain \xy\ positions for
all required objects on the plate.  In this section we outline our
plate solution approach which is sufficient to extract accurately
aligned spectra.  A full description is given in Bailer-Jones (1996).

\subsection{Evaluate Plate Centre}
A list of extraction targets for a given plate was drawn-up using the
plate codes which appear for each star in the MHD catalogue.  This
list was supplemented with PPM stars not listed in the MHD catalogue
on the basis of their \radec\ co-ordinates.  The coordinates of the
plate center, \cent, are only known to an accuracy of $\approx 1 \deg
$, corresponding to $>20$\% of the plate. Using this nominal centre,
the tangent plane projections, \xieta\ (the {\it standard
coordinates}), of the \radec\ positions of each star are obtained.
Once suitably scaled, the \xieta\ co-ordinates are the plate \1pos\
co-ordinates.  From the full list of extraction targets, a subset, the
$\Gamma_1$ spectra, is selected which will be used to define the first
plate solution. These spectra are those which are bright and
relatively isolated from other spectra, necessary to ensure their
unambiguous identification. We cannot use all spectra for forming the
plate solution at this stage on account of the poor nominal plate
centre.

\begin{figure}
\centerline{
\psfig{figure=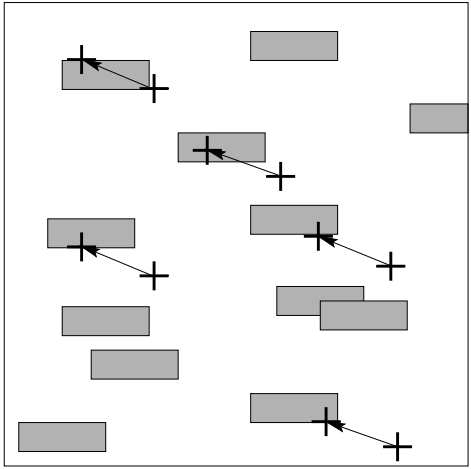,width=0.35\textwidth,angle=270}
}
\caption{An \xy\ offset is applied to the projected plate centre
in order to achieve a better superimposition of the object positions
with their spectra. This offset is determined visually.
The grey boxes are a schematic representation of the spectra, and the crosses
represent their initial and improved positions.}
\label{plt_2}
\end{figure}

The only interactive part of this reduction method is an iterative
procedure to improve the plate centre.  By displaying the \xy\
positions of the $\Gamma_1$ spectra over an image of the plate, the
$\Delta x$, $\Delta y$ shifts required to improve the match between
the spectra and positions are measured.  Using these offsets to move
the plate centre, the \xieta\ projections are recalculated and the
procedure repeated (Figure~\ref{plt_2}).  A good match can usually be
obtained in two iterations, taking only a couple of minutes. A highly
accurate plate centre is not required as the plate solutions include
constant terms which accommodate small linear offsets in $x$ and $y$.

\subsection{Marginal Sums and Cross-Correlation}
With perfect telescope optics and an exact plate centre, the \xieta\
co-ordinates would be sufficient to extract all spectra with known
\radec\ co-ordinates.  However, due to optical distortions, a plate
solution approach is needed. To achieve this, exact \xy\ plate
co-ordinates are required for the $\Gamma_1$ spectra.

\begin{figure}
\centerline{
\psfig{figure=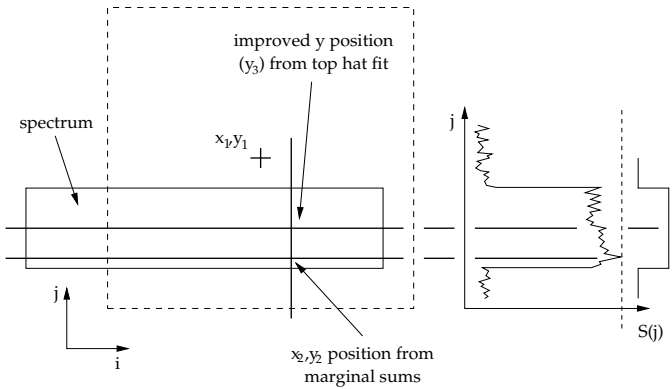,width=0.5\textwidth,angle=270}
}
\caption{Marginal sums in a region around the nominal \1pos\ position of
the spectrum yields the position \2pos.
The $y$-centre of the spectrum is found by 
fitting a top-hat across the spectral profile
function, $S(j)$, and taking the centre of the top-hat to
be the $y$-centre of the spectrum, $y_3$.}
\label{marg}
\end{figure}

Positions on the spectra are achieved using marginal sums, which
locates the brightest point within a rectangular box.
A box of size $800 \times 300$ pixels is centered on the \1pos\
position. The brightest point in this box is 
\begin{equation}
x_2 = \max_i S(i) = \max_i \left(\sum_{j=1}^{j=300} I_{i,j}\right)
\label{marg_x}
\end{equation}
and
\begin{equation}
y_2 = \max_j S(j) = \max_j \left(\sum_{i=1}^{i=800} I_{i,j}\right) \ \ ,
\label{y_marg}
\end{equation}
where $I_{i,j}$ is the number of (sky-subtracted) flux counts in pixel
$(i,j)$ (Figure~\ref{marg}).  As the nominal position, \1pos, of the
spectrum is uncertain, this box is considerably wider (300 pixels)
than the width of the spectrum (about 50 pixels).  The $y$-centre of
the spectrum ($y_3$) is then located by fitting a top-hat to $S(j)$.

The marginal sum $S(i)$ is a noisy version of the spectrum. Thus the
peak of $S(i)$, viz.\ $x_2$, differs for different spectral types (by
$\approx 50$ pixels), whereas we need to identify the position of a
common wavelength for all spectra. This is done by using the $x_2,
y_3$ positions to extract the $\Gamma_1$ spectra and then
cross-correlating them with templates (the extraction process is
described below).  The position of this cross-correlation peak ($x_3$)
corresponds to a common wavelength for all the
$\Gamma_1$ spectra.

\subsection{First Plate Solution}
The \3pos\ positions are used to solve the two-dimensional linear
plate solution equations
\begin{equation}
x_3 = a_0 + a_1\eta + a_2\xi
\label{x_3_eqn}
\end{equation}
and
\begin{equation}
y_3 = b_0 + b_1\eta + b_2\xi 
\label{y_3_eqn}
\end{equation}
for the 6 coefficients 
using Gauss--Jordan elimination (see, for example, 
Press et~al.\ 1992\nocite{press_92a}). 

\begin{figure}
\centerline{
\psfig{figure=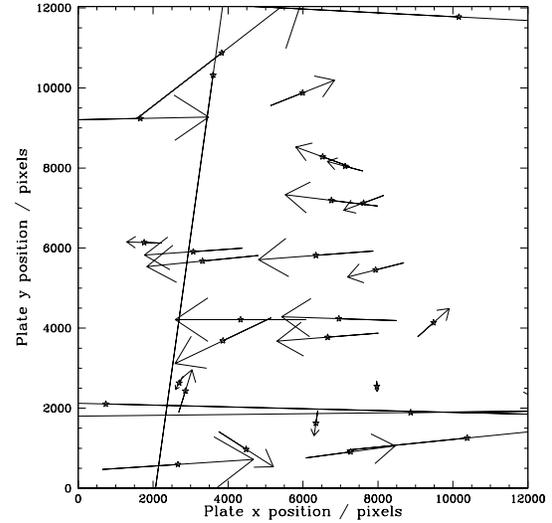,width=0.4\textwidth,angle=0}
}
\caption{Typical residuals after applying the first (2-D linear) plate
solution given by equations~\ref{x_3_eqn} and \ref{y_3_eqn}. The
length of each arrow is proportional to the magnitude of the solution
residual and the direction of
the arrow gives the relative sizes of the $x$ and $y$ errors. The
arrow at about (9500,4000) corresponds to a 2 pixel error.}
\label{resid1}
\end{figure}

Defining $x_3'$ as the values used to solve equation~\ref{x_3_eqn} and
$x_3$ as those obtained by applying the solution, the {\it solution
residual} is defined by $x_3 - x_3'$, and similarly for $y_3$.  The
equations were solved iteratively by rejecting, at each iteration (up
to a finite number of iterations), points which had residuals greater
than $3\tilde\sigma$, where $\tilde\sigma$ is the average of the
absolute value of the residuals.
(If the residuals are distributed as a Gaussian this would be
equivalent to $2.4\sigma$ clipping. 
The modulus error is less sensitive to outliers
than the RMS error and so gives a more stable error estimate
upon iteration.)
The final solution always had more than 25 objects,
which gave typical residuals of $\sigma_x
\approx 10$ pixels and $\sigma_y
\approx 1$ pixel. (These are not
the final errors: spectral alignment is
improved below.) 
Higher order solutions at this stage were found to be much less robust,
on account of the increased number of parameters.
Figure~\ref{resid1} shows
a typical example of the residuals plotted as a function of plate
position.  

\subsection{Spectral Extraction}
\begin{figure}
\centerline{
\psfig{figure=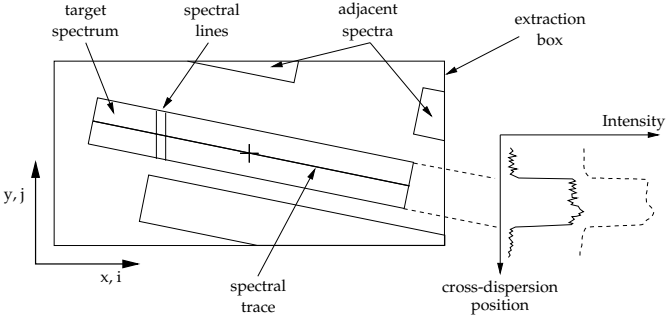,width=0.5\textwidth,angle=270}
}
\caption{Spectral extraction. The target spectrum is centred within
the extraction box. The spectrum is located, traced and extracted
using constraints on the position of the spectrum already determined.
Note that although the spectra are generally rotated relative to the
$x$-edge of the plate, the dispersion axis is still parallel to the
$x$-axis.  The solid line in the plot to the right is a schematic of
the profile at a point along the spectrum, and the dashed line the
corresponding aperture.}
\label{ext}
\end{figure}

Once solved, equations~\ref{x_3_eqn} and~\ref{y_3_eqn} give \xy\
positions for all spectra with known \radec\ co-ordinates.  An
extraction box of size $1020 \times 200$ pixels is placed at each
position and the {\sc apextract} routine from {\sc iraf}\footnote{{\sc
iraf} (Image Reduction and Analysis Facility) is distributed by the
National Optical Astronomical Observatory which is operated by the
Association of Universities for Research in Astronomy, Inc., under
contract to the National Science Foundation.} used to extract the
spectra.  Note that this extraction box is oversized in $y$ to ensure
that the ends of a rotated spectrum were included within the box, as
shown in Figure~\ref{ext}.  This rotation ($\sim 1\deg$) occurs
because the prism was not perfectly aligned relative to the East-West
axis of the telescope. Extraction is performed using apertures, based
on the optimal extraction algorithm first introduced by Hewett et~al.\
(1985) and subsequently generalized by Horne
(1986)\nocite{horne_86a}. The aperture is a model for the
cross-dispersion profile of the spectrum, with the optimum aperture at
each point determined by a maximum likelihood procedure (e.g.\ Irwin
1997). Because the location of the spectrum has been well-determined
in advance, it is guaranteed that the correct spectrum (as opposed to
an adjacent brighter spectrum) is traced and extracted.

Aperture fitting is done on sky subtracted pixels to increase the
dynamic range available for fitting. On account of the prism, the sky
background is grey and varies smoothly and slowly across the plate,
and was found to be uniform over the scale of a single spectrum
($\approx 0.4\deg$). The sky level is determined using an iteratively
$k$-$\sigma$ clipped median of all pixels in the extraction box.
Stellar pixels are preferentially removed with asymmetrical clipping
($k=1\sigma$ upper; $k=5\sigma$ lower).  The approach would be
invalid for very crowded regions where the pixels in the extraction
box are mostly stellar ones. However, in such cases there are also
large overlaps between the spectra making it very difficult to extract
the spectra anyway.

\subsection{Second Plate Solution}
\begin{figure}
\centerline{
\psfig{figure=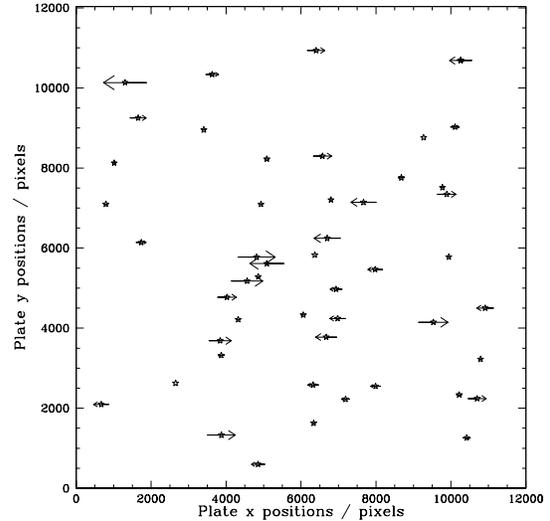,width=0.4\textwidth,angle=0}
}
\caption{Typical residuals after applying the second (1-D quadratic)
plate solution given by equation~\ref{sec_sol}.  The length of each
arrow is proportional to the magnitude of the solution residual and
the direction of the arrow gives the sign. The scale (length of
arrows) is the same as in Figure~\ref{resid1}.}
\label{resid2}
\end{figure}

We now have a set of one-dimensional extracted spectra aligned to a
precision of $\sigma_x \approx 10$ pixels.  This is improved upon by
locating a unique spectral feature (the \hb\ line) and using its
position to solve a second plate solution.  The \hb\ line is
suitable on account of being both strong and well-isolated from other
spectral lines in spectra earlier than about G5, thus easing
unambiguous identification.  A region is selected
around the expected position of the line, the continuum removed and
the spectrum inverted.  The \hb\ line is assumed to be the
strongest feature in this region which is at least $3\sigma$ above
the background. The mean of a Gaussian fitted to the line is
taken to be the position, $\Delta x$, of the \hb\ line relative
to $x_3$. 
 The spectra for which a line could be located (the $\Gamma_2$ spectra)
were used to solve the second plate solution
\begin{equation}
\Delta x = c_0 + c_1\eta + c_2\xi + c_3\eta\xi + c_4\eta^2 + c_5\xi^2 \ \ ,
\label{sec_sol}
\end{equation}
This was again solved iteratively using Gauss--Jordan elimination,
with approximately 50 spectra in the final solution.
Typical mean residuals for a given plate were $\sigma_{\Delta x}
\approx 1$ pixel, but a typical {\it median} residuals were $< 0.5$
pixels (Figure~\ref{resid2}).  Higher order solutions were found to be
less robust. Note that equation~\ref{sec_sol} assumes that
the prism dispersion is constant across the plate. This could
be relaxed using additional terms.

On account of the magnitude of these errors, alignment shifts
can be rounded-off to the nearest whole pixel. Alignment precisions
of better than $0.5$ pixels
require interpolation. One drawback of interpolation 
is that the noise in the resultant
spectrum is correlated between the pixels. This can be
problematic for subsequent analysis/classification algorithms.
Moreover, alignment precision for our spectra is limited by our ignorance
of the radial velocities of these stars. A typical line-of-sight
velocity of 40\kms gives a Doppler shift of $0.5$\,\AA\ at
4000\,\AA\, which corresponds to $\approx 0.5$ pixels.
Thus radial velocity variations across
the spectra limits alignment to no better than $0.5$ pixels.

In principle, cross-correlation with spectral standards could have
been used to align the spectra. However, the disadvantage of this
approach is that it requires that we know the approximate spectral
type in advance, so that the right standard can be selected.
Furthermore, a plate solution allows us to accurately extract faint
(low S/N) spectra which would give unreliable cross-correlations.

\section{Post-Extraction Processing}\label{post_ext}

The extracted spectra were cut to a final wavelength range of
3802--5186\,\AA, covered by 820 pixels.  This was dictated by the QE
of the telescope--prism--plate combination, and the need to retain at
least the region between the Ca II H\&K lines (at 3933.7\,\AA\ and
3968.5\,\AA) and \hb\ line (at 4861.3\,\AA) for use in the automated
classifiers.  The range was extended as far as there were still
spectral features at a reasonable S/N. The IIaO emulsion `cut-off' (where
the response drops to 50\% of the peak) occurs at 4900\,\AA,
although as can be seen from Figure~\ref{contp}, the drop-off in
response is slow. At the blue end, a blocking filter dramatically
reduces the QE below 3850\,\AA.

Most spectra were well-extracted and aligned.  In a few
cases---particularly for crowded plates, i.e.\ at low Galactic
latitudes---we discovered that some spectra
overlapped with neighbouring spectra.  These should ideally have been
deselected at the beginning of the reduction process based on their
proximity to other spectra, but they remained, presumably because 
the MHD catalogue (even supplemented with the PPM) is
not complete.  (Most of the plates were deliberately chosen to lie at
high Galactic latitudes to minimize crowding.)  A small fraction of
spectra were also deselected if they had unusually low S/N ratios,
possibly on account of a poor aperture fit during the spectral
extraction. A number of spectra were also lost due to overlap with the
edge of the plate. The total number of stars retained was 12,104 out
of a possible 15,820 spectra present on the 92 plates and listed in
the catalogues.

\section{Continuum Removal}\label{cont_rem}

Continuum-free spectra are required for many modes of spectral
analysis. For example, in stellar classification, although a genuine
stellar continuum is closely related to the effective temperature of a
star, the continuum received at a telescope's detector is often
distorted by interstellar reddening, atmospheric extinction and
instrumental effects. A particular problem is the non-linear (and
uncalibrated) response of the photographic emulsion.

There are many different ways in which a stellar continuum can be
removed, but not all are suitable or reliable.  One approach is to fit
a polynomial or non-linear spline to the spectrum and then subtract it
 from the spectrum. However, the high order polynomial usually needed
requires many data points for its definition and is therefore likely
to be distorted by spectral lines. One improvement is to fit the
continuum only in pre-defined `continuum windows' (regions which are
relatively line-free) (Zekl 1982)\nocite{zekl_82a}, although the
drawback here is that the approximate classification must be known in
advance, as the location of these windows depends on spectral type.
Another improvement is to fit the polynomial only to `high points' in
the spectrum, but this requires the distinction between continuum and
line features which can be very difficult for later-type stars.

Continuum removal is a process which removes all of the slowly
varying---low-frequency---information from a spectrum.  An attractive
approach is to take the Fourier transform of the spectrum, filter out
the low-frequency components (high-pass filter) and then
reverse-transform the spectrum back into wavelength space; this will
remove all slowly varying features.  The drawback of this Fourier
technique is that the broad spectral lines contribute to the
low-frequency components, so removing low frequencies alters some of
the line profiles and equivalent widths. LaSala \& Kurtz
(1985)\nocite{lasala_85a} improve upon this basic Fourier method by
defining a continuum by passing the Fourier-transformed spectrum
through a low-pass filter and Fourier-transforming back the result.
This gives a suitably smoothed version of the original spectrum.  The
original spectrum is then rectified by {\it dividing} it by this
continuum. This appears to give very reliable results for spectral
types earlier than M1, but the authors report that it ``fail[s]
catastrophically'' for later types and extreme emission line stars,
because in such cases the defined continuum can be negative in places.

We chose to use a combination of {\it median} and {\it boxcar
filtering} of a spectrum to obtain its continuum.  This is a
non-linear method which overcomes the shortcomings of linear methods
based on Fourier transforms.  The first process is to filter the
spectrum with a one-dimensional 
median filter.  Median filtering is performed by
replacing the flux in each pixel with the median value in a box of $M$
pixels centered on the pixel of interest.  The resulting `spectrum'
will not be very smooth, as it is composed of a sequence of 
flux values from the original spectrum which were generally 
non-adjacent. 
This `spectrum' is a non-linear transformation of the original
spectrum.  To smooth it, it is then boxcar filtered: This is like
median filtering except that each pixel is replaced with the {\it
mean} value in a box of size $N$. 
To obtain a reliable continuum at the ends of the spectrum,
a pseudo-spectrum is created beyond each end by reflecting
the spectrum about the end pixel. This gives better results than
simply truncating the filter size near the ends.
These combined filters produce a smooth continuum which is
subtracted from the original spectrum to give a line-only spectrum.
The sizes of the filter boxes depend on the scale over which the
spectrum shows variations. For our 820-pixel sized spectra, the values
$M$=101 and $N$=50 were found to be most suitable.  

\begin{figure}
\centerline{
\psfig{figure=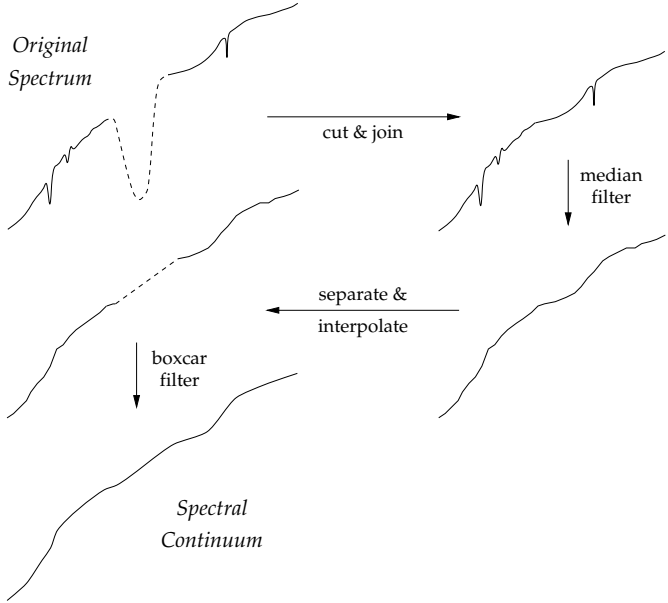,width=0.5\textwidth,angle=270}
}
\caption{Continuum evaluation by masking strong lines (schematic). 
The region of a strong
line is cut from the spectrum, the remaining spectrum joined up, and a
median filter passed across the spectrum. The resulting spectrum is
then split at the point where the spectral line was, and the spectrum
linearly interpolated across the gap. A linear boxcar filter is run across
this, resulting in the stellar continuum.}
\label{contrem}
\end{figure}

\begin{figure}
\centerline{
\psfig{figure=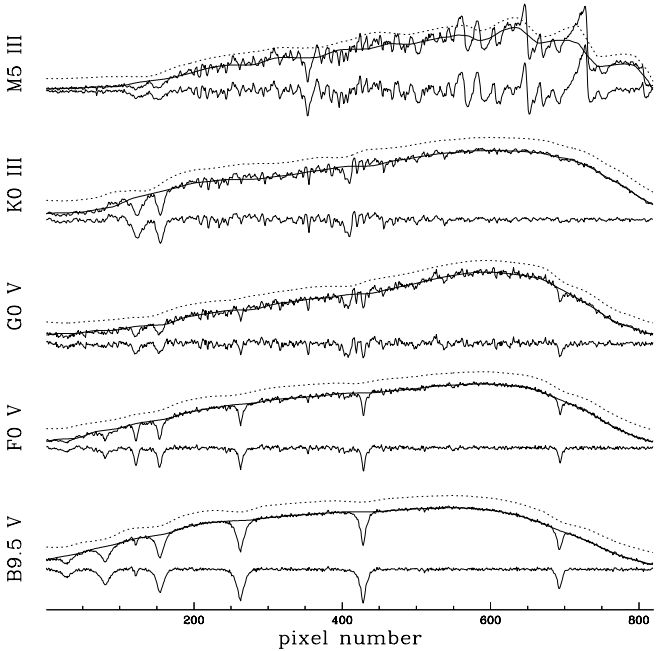,width=0.5\textwidth,angle=0}
}
\caption{Continuum fit and subtraction using median and 
boxcar filtering. For each spectral type the upper spectrum is the unrectified
spectrum, the solid line superimposed on it is the continuum obtained
using the masked filters, and the lower spectrum is the resultant
continuum-subtracted spectrum. The dashed line above each spectrum
shows the continuum obtained using unmasked filters.
Note that the `unmasked' continuum gives a poorer fit in the region
of broad lines.}
\label{contp}
\end{figure}

The continuum fits from this method are generally good, but are poor
in the regions of broad lines.  To overcome this problem, we masked
(cut out) the strong lines prior to median filtering, as shown in
Figure~\ref{contrem}. The masked and unmasked continua produced on a
range of spectral types are shown in Figure~\ref{contp}. It can be
seen that the masked continua are better near the strong lines,
particularly the hydrogen lines.  The large filter sizes of the
unmasked filtering reflected the width of the broad lines. With
masking, these sizes were reduced to $M$=51 and $N$=25.  The
wavelength coverages of the masked regions are shown in
Table~\ref{masked_tab}. 
\begin{table}
\begin{center}
\caption{Masked line regions in an improved median filtering.}
\begin{tabular}{ll} \hline
H + others      &  3811--3853\,\AA \\
H + Fe I        &  3864--3908\,\AA \\
Ca II H\&K    &  3924--3987\,\AA \\
H$\delta$       &  4078--4129\,\AA \\
CN G-band       &  4293--4320\,\AA \\
H$\gamma$       &  4325--4365\,\AA \\
H$\beta$        &  4837--4897\,\AA \\ \hline
\end{tabular}
\label{masked_tab}
\end{center}
\end{table}

Continuum fits at the redder ends of late-type stars are always poor:
the presence of many molecular bands makes the definition of a
`continuum' rather meaningless, so we can only remove low frequency
variations.  The main concern of continuum removal should be to
extract a continuum to `first order' in a {\it consistent} way, so as
to remove that continuum information which is not intrinsic to the
stellar spectrum, such as that produced by instrumental effects.
Provided this condition is met, the exact shape of the continuum which
is subtracted is not that important.  This is demonstrated by the
quality of the classifications we achieve with the reduced spectra
(Bailer-Jones et~al.\ 1997).

A combination of masked median filtering and linear filtering
generally gives better continuum fits than Fourier methods.  Any
Fourier continuum estimation 
method which involves filtering out the high frequency
components of the power spectrum is equivalent to `blurring' the
original spectrum by convolving it (in the wavelength space) with a
broad bell-shaped function. As such, the continuum will always be
distorted by the presence of broad lines or rapid changes in the
original continuum.  This convolution is a linear operation, which is
why Fourier methods are limited in the type of continua they give.
Median filtering, on the other hand, is a non-linear operation and can
therefore produce a better fit to the continuum. When followed up with
a linear filter (boxcar), a smooth continuum is obtained.  The
combined median/boxcar filter is also robust and consistent, in the
sense that it is not sensitive to data `spikes' (unlike linear
methods) and thus will give similar continua for similar spectral
types even in the presence of bogus spectral features.

\section{Summary}

\begin{figure}
\centerline{
\psfig{figure=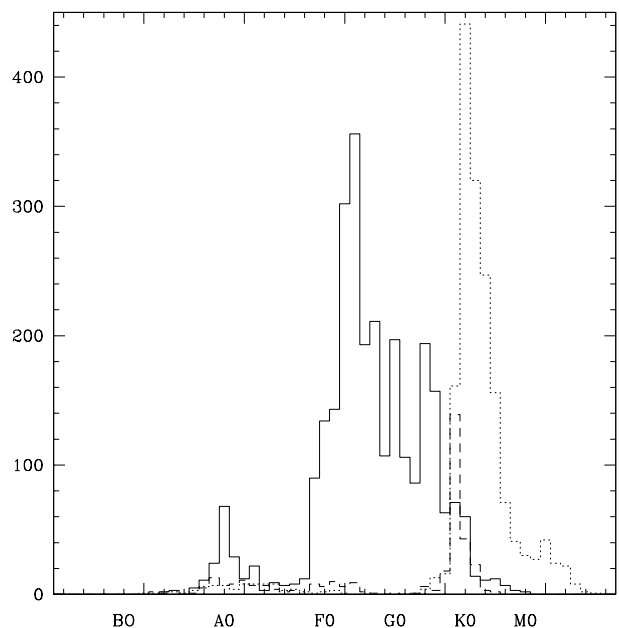,width=0.5\textwidth,angle=0}
}
\caption{Distribution of spectral types for each luminosity class.
The dotted line represent
giants (III), the dashed line subgiants (IV) and the solid line dwarfs (V).}
\label{dist_B}
\end{figure}

This paper has described a method for extracting spectra from
objective prism images.  The method has been developed for the
reduction of a set of photographic objective prism plates, but because
the spectral extraction and processing takes place entirely in
software using the complete digitized plate, it can equally well be
applied to CCD objective prism images.  The extraction process is
driven by a set of catalogue Right Ascension and Declination
positions, so a direct image of each field is not required. After an
initial interactive period taking one or two minutes, the subsequent
reduction is automatic, taking approximately one hour on a
modest-sized SUN Sparc IPX to process a single plate (i.e.\ extract
about 150 spectra).

The reduction method described in this paper has been used to extract
a set of over 12,000 high-quality spectra.  From this, a subset of
over 5,000 normal spectra was selected which had reliable
two-dimensional (spectral type and luminosity class) classifications
listed in the MHD catalogue.  The frequency distribution of the
various stellar classes in this set is shown in
Figure~\ref{dist_B}. This data set is used in accompanying papers to
produce automated systems for classifying and physically parametrizing
stellar spectra (Bailer-Jones et~al.\ 1997, 1998).

In the interests of extending spectral classification to more distant
stellar populations, spectra of stars fainter than B $\sim 12$ are
required. This could be achieved with a CCD objective prism survey.
Although the technique described can only extract objects with known
Right Ascension and Declination coordinates, the HST Guide Star
Catalogue (e.g.\ Lasker et al.\ 1990)\nocite{lasker_90a}, which lists
19 million objects brighter than 16$^{th}$ magnitude, could be used as
a driver for extraction. However, Bailer-Jones (unpublished, 1996) has
also modified the method to extract unwidened spectra from CCD
objective prism images in the absence of any coordinates, using an
algorithm to locate local flux peaks.  The method can be applied to
spectra at different spectral resolutions and wavelength coverages,
provided a suitable line exists for the second plate solution.

\section*{Acknowledgments}

We thank Nancy Houk for kindly loaning us her plate
material and an anonymous referee for useful comments.




\begin{thebibliography}{}
\bibliographystyle{plain}


\bibitem{bailerjones_96a}
Bailer-Jones C.A.L., 1996, PhD thesis, Univ.\ Cambridge

\bibitem{bailerjones_97a} 
Bailer-Jones C.A.L., Irwin M., Gilmore G.,
von Hippel T., 1997, MNRAS, 292, 157

\bibitem{bailerjones_98a} 
Bailer-Jones C.A.L., Irwin M., von Hippel T., 
von Hippel T., 1998, MNRAS, in press

\bibitem{clowes_84a}
Clowes R.G., Cooke J.A., Beard S.M., 1984, MNRAS, 207, 99

\bibitem{flynn_90a}
Flynn C., Morrison H.L., 1990, AJ, 100, 1181

\bibitem{hewett_85a}
Hewett P.C., Irwin M.J., Bunclark P., Bridgeland M.T., Kibblewhite E.J.,
He X.T., Smith M.G., 1985, MNRAS, 213, 971

\bibitem{horne_86a}
Horne K., 1986, PASP, 98, 609

\bibitem{houk_78a}
Houk N., 1978, University of Michigan Catalogue of Two-Dimensional Spectral
  Types for the HD Stars. Vol. 2: Declinations -53 to -40 degrees

\bibitem{houk_82a}
Houk N., 1982, University of Michigan Catalogue of Two-Dimensional Spectral
  Types for the HD Stars. Vol. 3: Declinations -40 to -26 degrees

\bibitem{houk_75a}
Houk N., Cowley A.P., 1975, University of Michigan Catalogue of
Two-Dimensional Spectral Types for the HD Stars.
Vol. 1: Declinations -90 to -53 degrees

\bibitem{houk_88a}
Houk N., Smith-Moore M., 1988, University of Michigan Catalogue of
Two-Dimensional Spectral Types for the HD Stars. Vol. 4:
Declinations -26 to -12 degrees

\bibitem{houk_94a}
Houk N., 1994, in Corbally C.J., Gray R.O.,
Garrison R.F., eds, Astronomical Society of the Pacific Conference
Series 60, The MK Process at 50 Years.  Astronomical Society of the
Pacific, San Francisco, p.\ 285

\bibitem{hagen_95a}
Hagen H.-J., Groote D., Engels D., Reimers D., 1995, A\&AS, 111, 195

\bibitem{irwin_97a}
Irwin M.J., 1997, in Espinosa J.M. ed., 
$7^{th}$ Canary Islands Winter School,
Instrumentation for Large Telescopes, in press

\bibitem{kibblewhite_84a}
Kibblewhite E.J., Bridgeland M.T., Bunclark P.S., Irwin M.J., 1984, in 
Kinglesmith D.A., ed, NASA-2317, Astronomical Microdensitometry
Conference. NASA, Washington D.C., p.\ 227

\bibitem{lasala_85a}
LaSala J., Kurtz M.J., 1985, PASP, 97, 605

\bibitem{lasker_90a}
Lasker B.M., Sturch C.R., McLean B.J., Russell J.L., 
Jenkner H., Shara M.M., 1990, AJ, 99, 2019

\bibitem{morgan_43a}
Morgan W.W, Keenan P.C., Kellman E., 1943,
An Atlas of Stellar Spectra with an Outline of Spectral
  Classification.
University of Chicago Press, Chicago

\bibitem{press_92a}
Press W.H., Teukolsky S.A., Vetterling W.T., Flannery B.P., 1992,
Numerical Recipes, 2nd edn.
Cambridge Univ.\ Press, Cambridge

\bibitem{roeser_91a}
R\"{o}ser S., Bastian U., 1991, PPM Star Catalogue, vols.~1 and 2.
Spektrum Akademischer Verlag, Heidelberg

\bibitem{vonhippel_94a}
von Hippel T., Storrie-Lombardi L., Storrie-Lombardi M.C., Irwin M., 
1994, MNRAS, 269, 97

\bibitem{wisotzki_96a}
Wisotzki L., Koehler T., Groote D., Reimers D., 1996, A\&AS, 115, 227

\bibitem{zekl_82a}
Zekl H., 1982, A\&A, 108, 380

\end{thebibliography}
\end{document}